\documentclass[twocolumn]{aastex631}

\shorttitle{AASTeX v6.3.1 Sample article}

\shortauthors{Jones et al.}

\graphicspath{{./}{figures/}}
\begin{document}

\UseRawInputEncoding

  

%

\author[0000-0001-7725-2546]{Evan Jones}
\affiliation{Physics and Astronomy Department, University of California, Los Angeles, 475 Portola Plaza, Los Angeles, CA 90095, USA}
\correspondingauthor{Evan Jones, Tuan Do}
\email{evan.jones@astro.ucla.edu, tdo@astro.ucla.edu}
\author[0000-0001-9554-6062]{Tuan Do}
\affiliation{Physics and Astronomy Department, University of California, Los Angeles, 475 Portola Plaza, Los Angeles, CA 90095, USA}
\author{Bernie Boscoe}
\affiliation{Physics and Astronomy Department, University of California, Los Angeles, 475 Portola Plaza, Los Angeles, CA 90095, USA}
  \affiliation{Computer Science Department, Occidental College, 1600 Campus Road, Los Angeles, California 90041, USA}
 \author[0000-0001-5436-8503]{Jack Singal}
 \affiliation{Physics Department, University of Richmond, 138 UR Drive, Richmond, VA 23173, USA}
\author{Yujie Wan}
\affiliation{Physics and Astronomy Department, University of California, Los Angeles, 475 Portola Plaza, Los Angeles, CA 90095, USA}
\author{Zooey Nguyen}
\affiliation{Physics and Astronomy Department, University of California, Los Angeles, 475 Portola Plaza, Los Angeles, CA 90095, USA}
\collaboration{6}{}

\title{Improving Photometric Redshift Estimation for Cosmology with LSST using Bayesian Neural Networks}

\begin{abstract}

We present results exploring the role that probabilistic deep learning models can play in cosmology from large-scale astronomical surveys through photometric redshift (photo-z) estimation. Photo-z uncertainty estimates are critical for the science goals of upcoming large-scale surveys such as LSST, however common machine learning methods typically provide only point estimates and lack uncertainties on predictions. We turn to Bayesian neural networks (BNNs) as a promising way to provide accurate predictions of redshift values with uncertainty estimates. We have compiled a galaxy data set from the Hyper Suprime-Cam Survey with grizy photometry, which is designed to be a smaller scale version of large surveys like LSST. We use this data set to investigate the performance of a neural network (NN) and a probabilistic BNN for photo-z estimation and evaluate their performance with respect to LSST photo-z science requirements. We also examine the utility of photo-z uncertainties as a means to reduce catastrophic outlier estimates. The BNN outputs the estimate in the form of a Gaussian probability distribution. We use the mean and standard deviation as the redshift estimate and uncertainty.  We find that the BNN can produce accurate uncertainties. Using a coverage test, we find excellent agreement with expectation -- 67.2$\%$ of galaxies between $0 < 2.5$ have 1-$\sigma$ uncertainties that cover the spectroscopic value. We also include a comparison to alternative machine learning models using the same data.  We find the BNN meets two out of three of the LSST photo-z science requirements in the range $0 < z < 2.5$.
\end{abstract}

\section{Introduction}

Dark matter and dark energy comprise $\sim 95\%$ of the energy density of the universe, but their natures are largely unknown. To investigate these entities, large-scale extragalactic surveys such as the Legacy Survey of Space and Time \citep[LSST -- e.g.][]{ivezic_lsst_2008} and Euclid \citep[e.g.][]{euclid_collaboration_euclid_2022} will soon provide observations of billions of galaxies. Cosmological probes of dark matter and dark energy aim to measure the structure and evolution of the universe, and thus rely in part on accurately and precisely measuring galaxy redshifts of hundreds of millions of galaxies with well-constrained uncertainties. Therefore the task of obtaining sufficiently accurate photometric redshift estimates and understanding the error properties of these estimates is a major challenge.

Spectroscopic redshift measurements are the most reliable method of obtaining redshift, but are too time consuming and therefore not a suitable solution for obtaining the number of redshifts required for cosmological measurements. Photometric redshift estimation (photo-z) can provide redshifts for billions of galaxies, however photo-z estimates are subject to significant systematic errors because the spectral information of a galaxy is sampled with only a limited number of imaging bands. These systematic errors can manifest as outlier predictions that are far from their true redshift, biases in the distribution of redshift predictions, and large scatter in redshift predictions \citep[e.g.][]{newman_photometric_2022}. These systematics strongly affect science goals such as weak lensing inferences of cosmological parameters since photo-z uncertainties will be propagated into models constraining cosmological quantities. Any photo-z model developed for the potential application to these science missions must produce uncertainties on photo-z predictions.
  
 According to the LSST Science Requirements Document (SRD)\footnote{https://docushare.lsstcorp.org/docushare/dsweb/Get/LPM-17}, sufficiently accurate photo-z estimates for $\sim$ four billion galaxies are required to meet the LSST science goals for their main cosmological sample. Specifically, for the $i < 25$ flux-limited galaxy sample measured by LSST, one must achieve

\begin{itemize}
  \item number of galaxies $\approx 10^7$ 
  \item rms error $<$ 0.2 (Equation 3 in Table 2)
  \item bias $<$ 0.003 (Equation 4 in Table 2)
  \item 3$\sigma$ catastrophic outliers $<$ 10$\%$ total sample (Equation 2 in Table 2)
\end{itemize}

Currently, no published model satisfies the LSST photo-z science requirements up to $z = 3$ \citep{tanaka_photometric_2018-1, schuldt_photometric_2020-1, schmidt_evaluation_2020-1}. Additionally, methods for rejecting the majority of outliers and characterizing their effects on the predictions must be developed \citep{ivezic_lsst_2018}. Beyond the LSST metrics stated in the SRD, we consider additional probabilistic metrics for quantifying the quality of uncertainty estimates  \citep[see Table 2 --][]{malz_how_2020, schmidt_evaluation_2020-1, jones_photometric_2022}. The requirement thresholds for the probabilistic metrics are not as well quantified at this time as those for point metrics, but they allow us to compare the performance between different probabilistic models evaluated on the same data. Techniques for identifying photo-z outlier predictions in machine learning models have been investigated in \citet{jones_tests_2020, wyatt_outlier_2020}, and \citet{singal_machine_2022}.

Photo-z estimation techniques have traditionally been divided into two main approaches. Template fitting methods, such as Lephare \citep{ilbert_accurate_2006, arnouts_measuring_1999}, Mizuki \citep{nishizawa_photometric_2020}, and Bayesian Photometric Redshift \citep[BPZ --][]{benitez_bayesian_2000}, involve correlating the observed band photometry with model galaxy spectra and redshift, and possibly other model properties. Machine learning methods, such as artificial neural networks \citep[e.g. ANNZ --][]{collister_annz_2004}, boosted decision trees \citep[e.g. ARBORz --][]{gerdes_arborz_2010}, regression trees / random forests \citep{carrasco_kind_tpz_2013}, support vector machines \citep[SVMs -- e.g.][]{wadadekar_estimating_2005, jones_analysis_2017,jones_tests_2020}), a Direct Empirical Photometric method \citep[DEmP --][]{tanaka_photometric_2018}, and others develop a mapping from input parameters to redshift with a training set of data in which the actual spectroscopic redshifts are known, then apply the mappings to data for which the redshifts are to be estimated. Both have their drawbacks -- template fitting methods require assumptions about intrinsic galaxy spectra or their redshift evolution, and empirical methods require the training set and evaluation set to significantly overlap in parameter space. As machine learning approaches for photo-z estimation have increased in capability and larger data sets have been observed over the past decade, galaxy images can be effectively used as inputs to utilize morphological information for photo-z estimation, unlike template-fitting approaches. 

\begin{figure*}[!th]
\centering
{\includegraphics[width=6in]{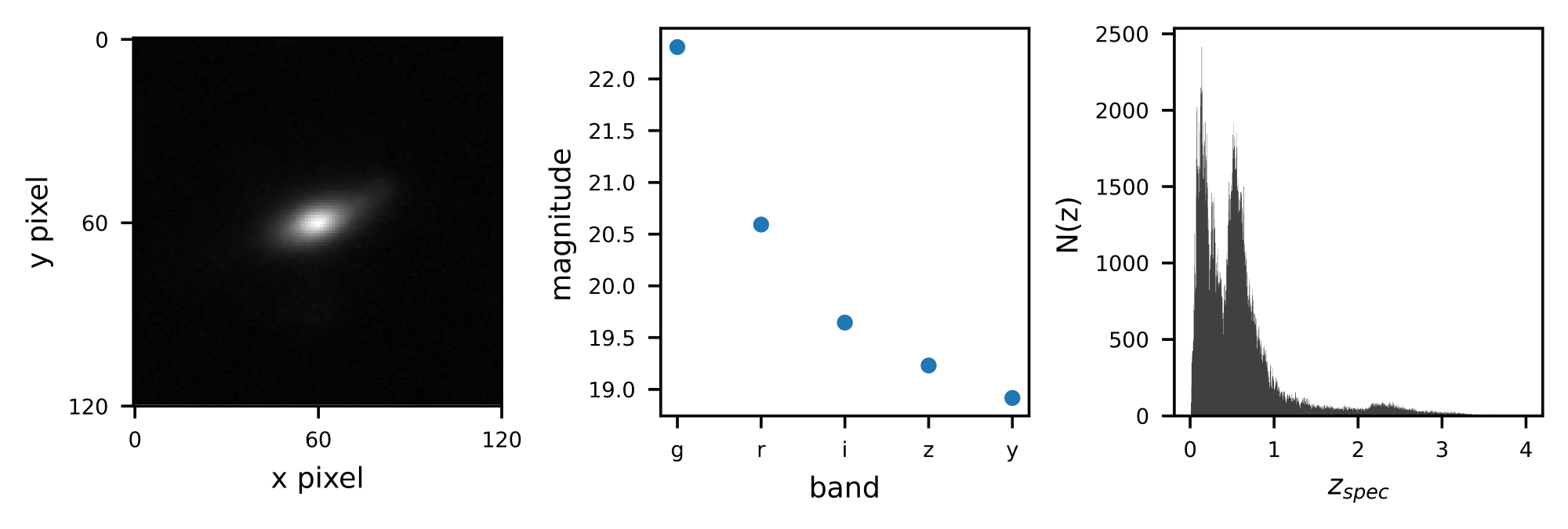}}
\label{z_rems}
\caption{Left: Example of a galaxy ($z$ = 0.48) image in the $i$-band. Middle: five-band photometry for the same galaxy. Right: $N(z)$ distribution for the data set discussed in \S \ref{data} For the photo-z determinations in this work we use training, validation, and  testing sets consisting of 229,120, 28,640, and 28,640 galaxies respectively.}
\label{fig1}
\end{figure*}

There have been a number of works studying the application of neural networks for photo-z estimation \citep{firth_estimating_2003, collister_annz_2004, singal_efficacy_2011}, while probabilistic NN techniques have had limited investigations until recently \citep{sadeh_annz2_2016, pasquet_photometric_2019, schuldt_photometric_2020-1, zhou_photometric_2022}. Bayesian neural networks (BNN), a type of probabilistic NN \citep{jospin_hands-bayesian_2020} are a promising approach that has not been well explored. Probabilistic neural networks, conceptualized in the 1990s \citep{specht_probabilistic_1990}, have previously been limited in their ability to process the size of data required for performing photo-z estimation for large-scale surveys, because of the complexity of their computation. However, recent breakthroughs in conceptual understanding and computational capabilities \citep[e.g.][]{filos_systematic_2019, dusenberry_analyzing_2020} now make probabilistic deep learning possible for cosmology. Probabilistic deep learning with a BNN has many advantages compared to traditional neural networks, including better uncertainty representations, better point predictions, and offers better interpretability of neural networks because they can be viewed through the lens of probability theory. In this way we can draw upon decades of development in Bayesian inference analyses.

We have three goals in this work: (1) develop a probabilistic ML model that can produce robust uncertainties for photometric redshifts, (2) assess the model with respect to LSST requirements and alternative photo-z estimation methods, and (3) investigate the use of photo-z uncertainties to identify likely outliers in photometric redshift predictions. For the analysis in this work, we have created the largest publicly available machine-learning-ready galaxy image data of $\sim 300k$ galaxies from the Hyper Suprime-Cam survey containing five-band photometric images and known spectroscopic redshifts from $0 < z < 4$. This data will be released in Do et al. 2024 (in prep). In \S 2 we discuss the data and network architecture. In \S 3 we state the results. In \S 4 and \S 5 we provide a discussion and conclusion.

\section{Data and Methods}

\subsection{Data: Galaxy observations}

For the analysis in this work we compile a data set intended to approximate the data produced by future large-scale deep surveys for photo-z estimation \citep{the_lsst_dark_energy_science_collaboration_lsst_2021}.  We use the Hyper-Suprime Cam (HSC) Public Data Release 2 \citep{aihara_second_2019}, which is designed to reach similar depths as LSST but over a smaller portion of the sky. We choose the HSC survey because it mimics LSST in photometry and depth. Including photometry in infrared bands would improve photo-z estimates, but since LSST will provide observations in only optical bands \citep{ivezic_lsst_2008}, we will restrict our analysis to optical bands only. HSC is a wide-field optical camera with a FOV of 1.8 deg$^2$ on the Subaru Telescope. HSC PDR2 surveys more than 300 deg$^2$ in five optical filters ($grizy$). The median seeing in the i-band is 0.6''. This data set is presented in more detail in Do et al. 2024, in prep.

The final data set used in the analyses of this paper consists of $\sim$ 300k galaxies with 5-band grizy photometry and spectroscopic redshifts. Fig. 1 contains the N(z) distribution for the dataset and Fig. 2 contains grizy images for three example HSC galaxies. Spectro-zs were obtained by crossmatching galaxy photometry from HSC with the HSC collection of publicly available spectroscopic redshifts using galaxy sky positions (d $< 1^{\prime\prime}$) in \cite{lilly_zcosmos_2009}, \cite{bradshaw_high-velocity_2013}, \cite{mclure_sizes_2012}, \cite{skelton_3d-hst_2014}, \cite{momcheva_3d-hst_2016}, \cite{le_fevre_vimos_2013}, \cite{garilli_vimos_2014}, \cite{liske_galaxy_2015} \cite{davis_science_2003}, \cite{newman_deep2_2013}, \cite{coil_prism_2011}, \cite{cool_prism_2013}. We used data quality cuts similar to \cite{nishizawa_photometric_2020} and \cite{schuldt_photometric_2021} (see Table 1 and Do et al. 2024 (in prep) for a full list), which are intended to remove outlier photometric measurements and poorly measured spectroscopic redshifts. We also required detections in each band. The spectroscopic redshift values are treated as the ground truth for training and evaluation. In total, the data consists of  286,401 galaxies with broad-band grizy photometry and known spectroscopic redshifts.  Our galaxy sample extends from $0.01 < z < 4$, however the majority of the sample lies between redshift of 0.01 and 2.5 with peaks at $z ~ 0.3$ and $z ~ 0.6$ (see N(z) in Fig. \ref{fig1}). We use 80\% of the galaxies for training, 10\% for validation, and 10\% for testing. The data used for training is available\footnote{\url{https://zenodo.org/records/5528827}} from \citep{jones_2021_5528827}. This dataset includes the photometry and spectroscopic redshifts. A future release will also include images (Do et al. in prep.). 
\begin{figure*}[!tbh]
  \centering
  \begin{minipage}[b]{\textwidth}
    \includegraphics[width=\textwidth]{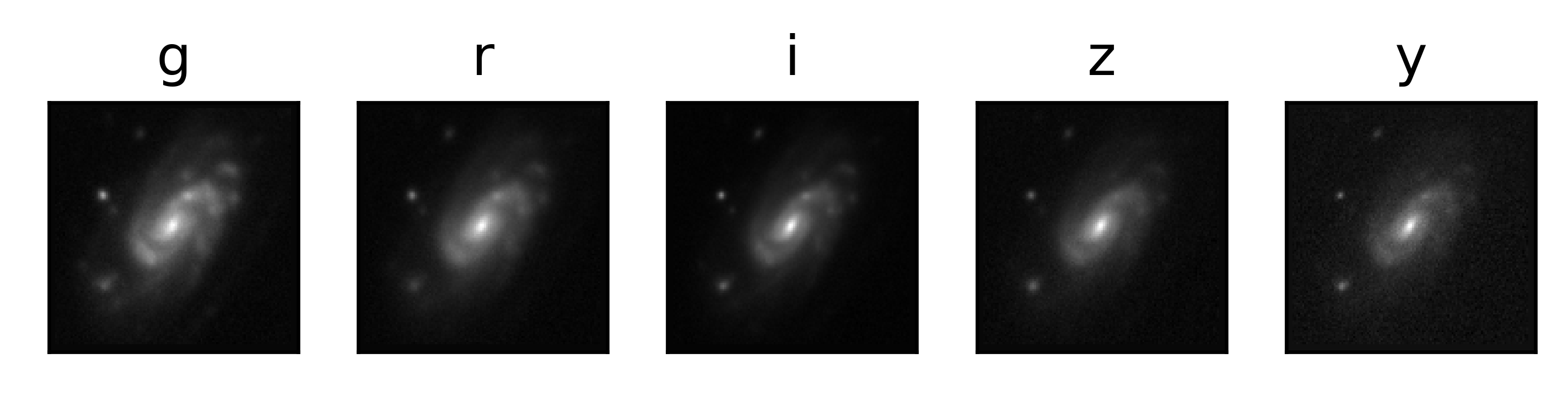}
  \end{minipage}
  \hfill
  \begin{minipage}[b]{\textwidth}
    \includegraphics[width=\textwidth]{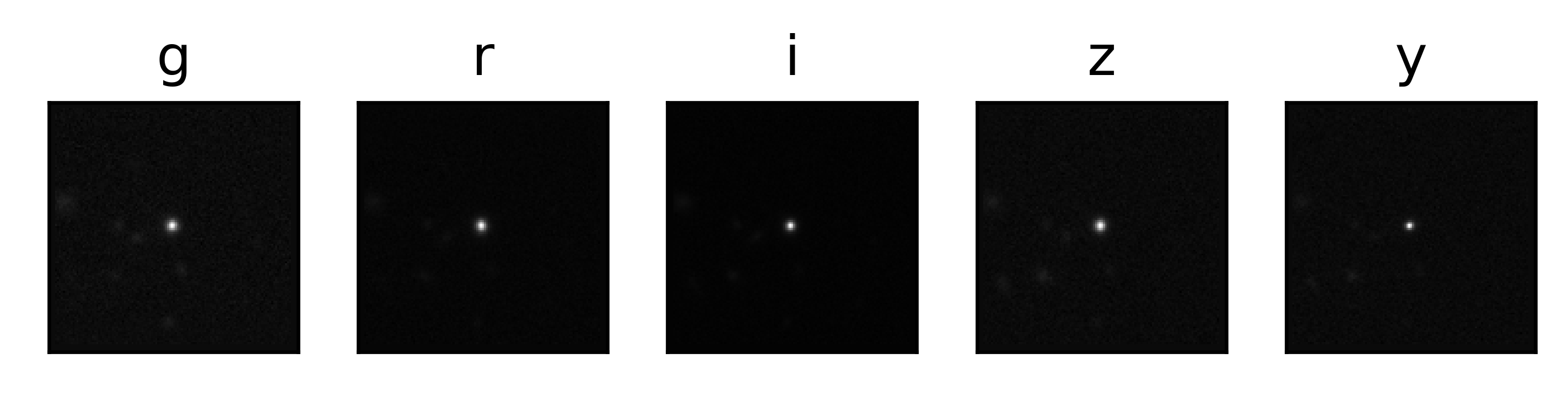}
  \end{minipage}
  \hfill
  \begin{minipage}[b]{\textwidth}
    \includegraphics[width=\textwidth]{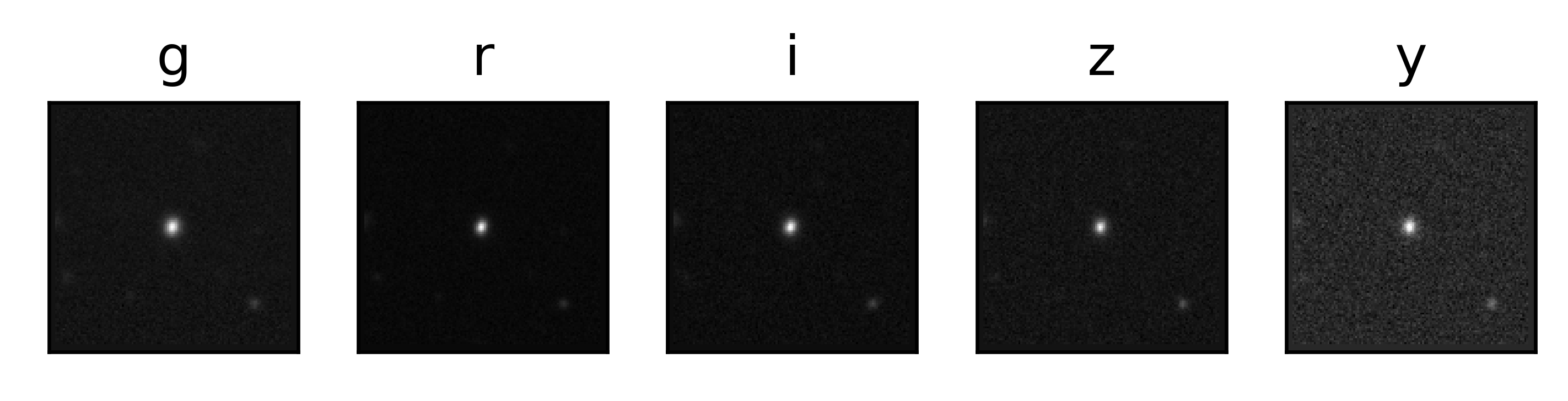}
  \end{minipage}
  \hfill
 \caption{Example HSC galaxy images for the data set used in this work with $grizy$ photometry for a low redshift galaxy at $z = 0.05$ (TOP), and a high redshift galaxy at $z = 3.92$ (MIDDLE), and another low redshift galaxy at $z = 0.14$ (BOTTOM). The similarity between the high redshift galaxy and the bottom low redshift galaxy highlights the difficulty of photo-z estimation.}
 \label{fig2}
\end{figure*}
\begin{table*} 
\caption{Quality cuts used to construct the data set.}
\centering
\begin{tabular}{lr}
\hline
\hline
photometry cuts & $z_{spec}$ cuts\\
\hline
\hline
\text{grizy\_ cmodel\_flux\_flag} = False & $z > 0$\\
\text{grizy\_ pixelflags\_edge} = False & $z \ne 9.9999$\\
\text{grizy\_ pixelflags\_interpolatedcenter} = False & $0 < z_{err} < 1$\\
\text{grizy\_ pixelflags\_saturatedcenter} = False & \text{unique galaxy object ID}\\
\text{grizy\_ pixelflags\_crcenter} = False & \text{specz\_flag\_homogeneous = True}\\
\text{grizy\_ pixelflags\_bad} = False\\
\text{grizy\_ sdsscentroid\_flag} = False\\
\hline
\end{tabular}
\end{table*}
\label{data}

\subsection{Network architectures}
\begin{figure*}[!tbh]
  \centering
  \begin{minipage}[b]{0.30\textwidth}
    \includegraphics[width=\textwidth]{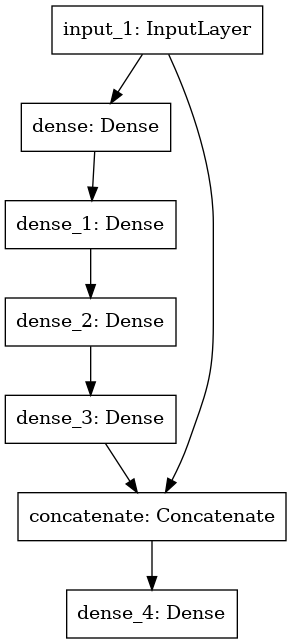}
  \end{minipage}
  \hfill
  \begin{minipage}[b]{0.45\textwidth}
    \includegraphics[width=\textwidth]{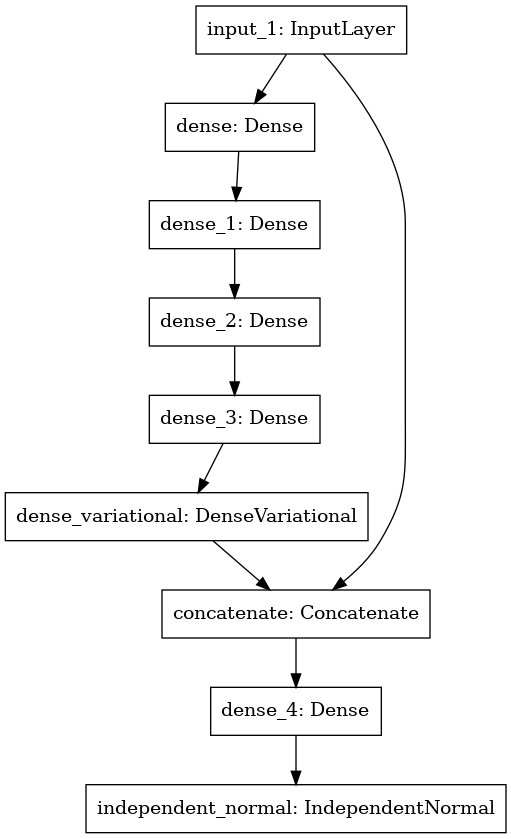}
  \end{minipage}
 \caption{Left: NN architecture. Right: BNN architecture. The inputs for both networks are five-band photometry in the $g$,$r$,$i$,$z$,$y$ filters. The output for the NN is a single point photo-z estimate while the output for the BNN is a photo-z PDF, which we sample to obtain a photo-z estimate. We assume Gaussianity in the creation of the photo-z PDF, so a photo-z uncertainty is produced by the standard deviation of the PDF.}
 \label{fig3}
\end{figure*}

We built two neural networks for this work -- one is a fully connected neural network that produces single-valued redshift predictions and one is a Bayesian neural network that outputs Gaussian probability distributions. The NN and BNN models are visualized in Fig. \ref{fig3}. Both the NN and the BNN are implemented in TensorFlow \citep{abadi_tensorflow_2016} and have five input nodes for the five-band $grizy$ photometry. We performed a parameter grid search to optimize for free parameters, such as the number of epochs, number of layers, number of nodes per layer, learning rate, loss function, activation function, and optimizer. Both the NN and BNN used for the final analysis in this work contain four hidden layers with 200 nodes per layer and utilize a rectified linear activation function. The networks also have a skip connection between the input nodes and the final layer. The NN has an output node to produce a single point estimate photo-z prediction while the BNN has a final output node that produces a mean and standard deviation assuming a Gaussian distribution for each photo-z prediction. For the BNN we use a negative log likelihood loss function with RMS error as the metric. We choose the negative log-likelihood loss function for the BNN because it has been shown to be more effective than MAE for probabilistic NNs \citep{lakshminarayanan_simple_2016}. The NN uses a mean absolute error loss function, and we also consider a custom loss function \citep{nishizawa_photometric_2020} defined in equation 6 of Table 2. The NN and BNN use the Adam optimizer and have learning rates of 0.0005 and 0.001, respectively. We train using an AMD Ryzen Threadripper PRO 3955WX with 16-Cores and NVIDIA RTX A6000 GPU. Training and evaluation runtimes are typically under 30 minutes. 

\subsection{Other ML models}
We use three other common ML models in order to compare to the neural network performance: (1) a support SVM classification model, (2) a random forest regression (RF) model, and (3) a gradient boosted tree regression model. For RF models we utilize the Scikit-Learn implementation \citep{2011JMLR...12.2825P} \footnote{https://scikit-learn.org/stable/modules/classes.html}. For the SVM we use SPIDERz \citep{jones_analysis_2017, jones_tests_2020}, which implements support vector classification on classes of redshift bins of width $z = 0.1$ spanning $0 < z < 4$. The RF model uses the RandomForestRegressor package to produce photo-z estimates. We also use the default hyperparameters with the RF, with the exception of using 200 trees in the forest. We use the XGboost software package for the gradient boosted tree model (the XGBRegressor library) with default hyperparameters. We perform a broad hyperparameter grid search for each model, however the performance boost over default parameters is not significant.

\label{network_architectures}

\label{architectures}

\section{Photo-z metrics}
Photo-z uncertainties are propagated to measurement uncertainties on dark matter and dark energy. Therefore, our choice of metrics to evaluate the photo-z determinations in this work include chiefly the photo-z metrics used in the LSST science requirements document (RMS error (Eq. 3), Bias (Eq. 4), and $3 \sigma$ Outliers (Eq. 7)) which are calculated to provide the necessary precision to constrain important cosmological quantities.  Specifically, for the purpose of constraining dark matter and dark energy we require photo-z RMS error ($< 0.2$), Bias ($< 0.003$), and $3 \sigma$ Outliers ($< 10\%$). In addition, we also include in our analysis a number of point metrics that are commonly used in the photo-z literature (Outlier (Eq. 1), Catastrophic Outlier (Eq. 2), Scatter (Eq. 5), and Loss (Eq. 6)) for the purpose of comparison to other models, as well as additional probabilistic metrics to evaluate the photo-z uncertainties produced by the BNN.  To measure model performance we evaluate predictions using the metrics in Table 2, which are separated into non-probabilistic and probabilistic categories. These metrics describe different ways to characterize the photometric redshift performance averaged over all predictions.  Ideally, photo-z measurements should be accurate out to the redshift limit of LSST observations (~z = 3.4 is where galaxies begin dropping out of the $g$ band), however the main redshift range of focus is $0.3 < z < 3.0$. In this redshift range, LSST aims to measure the comoving distance as a function of redshift to an accuracy of 1-2$\%$. In order to achieve this goal, LSST must obtain (1) a sufficiently large sample of galaxies ($\sim$ four billion) and (2) sufficiently accurate photo-z measurements for these galaxies as defined by the aforementioned requirements.  In addition to meeting photo-z science requirements, the LSST team also requires `methods for rejecting the majority of those outliers, and for characterizing their effects on the sample'.

We note that science missions of LSST and Euclid for which photo-zs are necessary divide the redshift ranges of interest into several discrete tomographic redshift bins ($0 < z < 1.5$ divided into four bins of $z = 0.3$ in weak lensing analyses).  The photo-z science requirements must be achieved on average throughout each tomographic redshift bin, rather than on average throughout the entire sample. This means that a full evaluation of a particular photo-z method must include an evaluation of important metrics as a function of redshift, rather than averaging across the entire photo-z sample. This distinction is particularly important for evaluating model performance of high redshift regions ($z > 1.0)$, which contain significantly fewer galaxies than low redshift regions (see Fig. \ref{fig1}), and are thus more challenging for any photo-z method to accurately produce photo-zs.

\subsubsection{Point Metrics}
We use the conventional definition for photometric redshift outliers and catastrophic outliers in Eqs. 1 and 2, where $z_{phot}$ and $z_{spec}$ are the estimated photo-z and actual (spectroscopically determined) redshift of the galaxy. The RMS photo-z error is given by a standard definition in Eq. 3, where $n_{gals}$ is the number of galaxies in the evaluation testing set and $\Sigma_{gals}$ represents a sum over those galaxies. Bias and scatter are defined in Eqs. 4 and 5. We follow \citet{tanaka_photometric_2018} and define a loss function in Eq. 6 to characterize the point estimate photo-z accuracy with a single number, where we use $\gamma = 0.15$.

\subsubsection{Probability metrics}

We propose {\it coverage} as a key metric for assessing the performance of the BNN (see Eq. 8). Coverage is typically used to assess whether confidence intervals are accurate. In this case, we define coverage as the fraction of galaxies that have a spectro-z within their 68$\%$ confidence interval. Ideally, 68$\%$ of evaluated galaxies should have true spectro-zs within their 68$\%$ confidence interval. If the coverage is over 68\%, then the estimated uncertainties are on average too large. Similarly, if the cover is below 68\%, the estimated uncertainties are on average too small. 

\label{metrics}
\begin{table*}
\centering
\caption{Metrics used to assess model performance.}
\begin{tabular}{lrrrr}
\hline
\hline
Point Metrics & & Probabilistic Metrics\\
\hline
\hline
Outlier & $O: {{\vert z_{phot}-z_{spec} \vert} \over {1+z_{spec}}} > .15$ (1) &  $3\sigma$ Outlier: &  $ {\vert z_{phot}-z_{spec} \vert} > 3 z_{\sigma}$ (7)\\
Catastrophic Outlier & $O_c: \vert z_{phot}-z_{spec} \vert > 1.0$ (2)\\
RMS error & $  \sqrt { {{1} \over {n_{gals}}} \Sigma_{gals} \left( {{ z_{phot}-z_{spec} } \over {1+z_{spec}}} \right) ^2 } $ (3) &  Coverage & $ \displaystyle\sum_i^{n_{gals}} \frac{(\bar{z}_{pdf,i} - z_{spec,i}) < z_{\sigma,i}}{n_{gals}} $ (8)\\ Bias & $b = {{ z_{phot}-z_{spec} } \over {1+z_{spec}}}$ (4)
  \\
Scatter &   Median($|\Delta z - $Median$(\Delta z_i)|)$ (5) &  PIT: & $\int_{-\infty}^{z_{spec}} p(z)dz$ (9)  \\
Loss & $  L(\Delta z) = 1 - \frac{1}{1+(\frac{\Delta z}{\gamma})^2} $ (6)\\

\hline
\end{tabular}
\end{table*}

Error in the bulk photo-z distribution width for the evaluation set can be difficult to distinguish between uncertainties associated with galaxy bias or uncertainties in the mean redshift of photo-z tomographic bins. The Probability Integral Transform (PIT) is a photo-z metric that can detect systematic error in the photo-z distribution width for galaxy samples with known spectroscopic redshifts \citep{malz_how_2020, malz_how_2021}. The PIT value for a single galaxy is defined in Eq. 9 in Table 2, where $p(z)$ is the predicted photo-z PDF. A histogram of PIT values for a galaxy sample should be uniform for an accurate collection of $p(z)$ samples. Ideally, the PIT histogram is flat across all redshift bins. If the PIT histogram peaks at the center, the $p(z)$ collection is too broad. If the PIT histogram peaks at high and low PIT values, the $p(z)$ samples are too narrow.  For a comparison of several probabilistic photo-z methods, see \citet{schmidt_evaluation_2020}.

\subsection{Leveraging BNN for Outlier Identification}

\begin{figure*}[!tbh]
  \centering
  \begin{minipage}[b]{0.45\textwidth}
    \includegraphics[width=\textwidth]{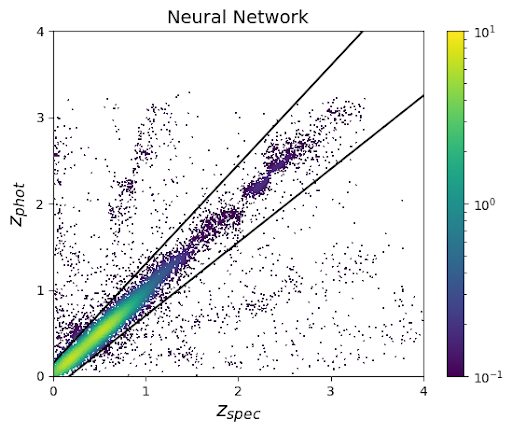}\\
    \includegraphics[width=\textwidth]{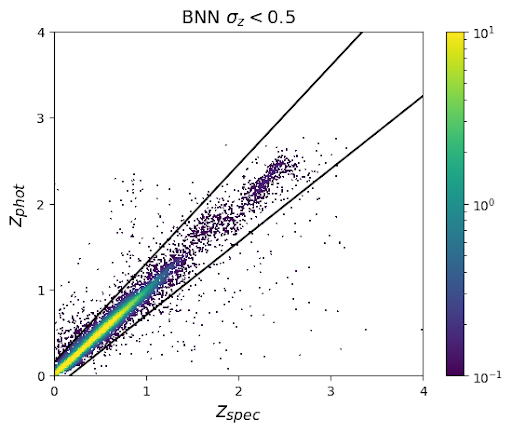}\\
  \end{minipage}
  \hfill
  \begin{minipage}[b]{0.45\textwidth}
    \includegraphics[width=\textwidth]{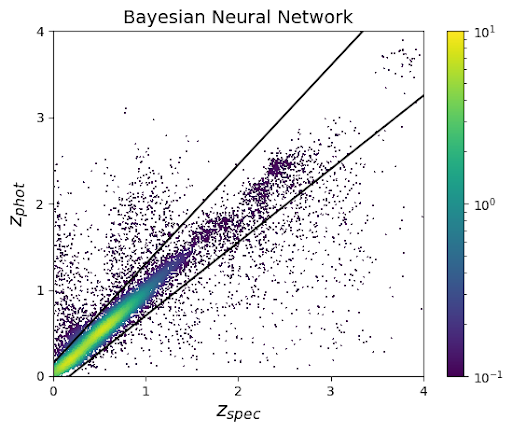}\\
    \includegraphics[width=\textwidth]{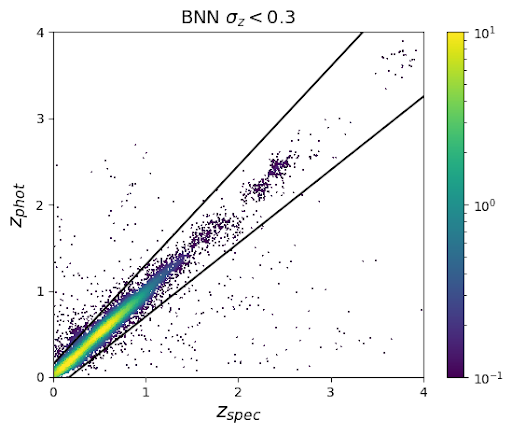}\\
  \end{minipage}
 \caption{Visualization of NN (top left) and BNN (top right) performance compared to the BNN with outlier removal criteria examples ${\sigma}_z$ = 0.5 (bottom left) and ${\sigma}_z$ = 0.3  (bottom right).}
 \label{fig4}
\end{figure*}

We propose a method for utilizing the photo-z uncertainties $z_{\sigma}$ produced by the BNN to preemptively flag photo-z predictions with high uncertainties as potential poor predictions. The method is simple: all galaxies with a photo-z uncertainty greater than the specified $z_\sigma$ cutoff value are flagged as potential outlier or catastrophic outlier candidates and removed from the evaluation sample. Figs. \ref{fig4}, \ref{fig5}, and \ref{fig6} depict performance improvements with example ${\sigma}_z$ removal values for a variety of performance metrics including the LSST photo-z requirements. An acceptable balance needs to be achieved between the number of galaxies correctly flagged as poor predictions versus the number of non-outlier galaxies removed for a given $z_{\sigma}$ cutoff value.  Other outlier removal strategies have previously been explored in \citet{jones_tests_2020}, \citet{wyatt_outlier_2020}, and \citet{singal_machine_2022}.

With the data used in this work we find a significant reduction in the number of catastrophic outliers and outliers by sacrificing a minimal number of non-outlier predictions; for example, we find that by removing all galaxies in the evaluation sample with a photo-z uncertainty ${\sigma}_z > 0.3$, the RMS error was reduced by 57.6$\%$, outliers were reduced by 70.1$\%$, and catastrophic outliers were reduced by $80.43\%$ -- at the cost of removing only 11$\%$ of the evaluation set. See Fig. \ref{fig9} for the N(z) distribution of removed galaxies for example cases of ${\sigma}_z > 0.3$ and ${\sigma}_z > 0.5$.

\section{Results}
\label{results}

The BNN generally satisfies LSST photo-z science requirements in the range of $0.3 < z < 1.5$ (redshift range for weak lensing analyses -- see Fig. \ref{fig8}) and performs as well or better than the 6 common alternative methods investigated in this study (see Table 3 and Figs. \ref{fig7} and \ref{fig8}). We compare the BNN and NN to a support vector machine \citep{cortes_support-vector_1995}, a random forest \citep{breiman_random_2001}, and a gradient boosting model, XGBoost \citep{chen_xgboost_2016}, using the same data discussed in \S \ref{data}. We also form a comparison to photometric redshift predictions measurements from the HSC team \citep{nishizawa_photometric_2020}, which used the template-fitting model Mizuki and empirical method DEmP \citep{ hsieh_estimating_2014}. To form a comparison, we relied on the photometric redshifts produced by the HSC team. Mizuki and DEmP were trained and evaluated on a slightly larger data set of 300k galaxies by \citep{nishizawa_photometric_2020}, but the majority of galaxies overlap with the data set introduced in this work. We crossmatched the HSC data set with the object IDs of our data to obtain a pre-evaluated sample of $\sim$60 thousand galaxies. Another photo-z investigation performed by \citet{schuldt_photometric_2021} utilized HSC imaging data and obtained a precision of $\Delta_z = |z_{phot} - z_{spec}| = 0.12$ with a convolutional neural network averaged over all galaxies in the redshift range $0 < z < 4$. We obtain $\Delta_z = 0.0031$ for the NN and $\Delta_z = 0.0032$ for the BNN averaged over all galaxies in our data set in this range. We note that a perfect comparison between photo-z models requires identical training, validation, and evaluation data sets. While the photo-z models from \cite{schuldt_photometric_2020} and the HSC team compared in this work utilized largely the same data that was used in this work, there are some differences between their data and the data used in this investigation, which introduces additional uncertainty in the comparison made between results. 

We note that training a Bayesian neural network does not deterministically produce weights on the same data. The weights in the variational layers are sampled from a Gaussian distribution.  The results presented here are representative of a typical training run with the BNN model presented in this work. However, there can be variations of several percents in outlier rates and other metrics depending on the training run. There can also be variations in the final loss achieved at the end of training. We find that the accuracy of a particular training run is correlated to the final loss value. 

\subsection{Using BNN Uncertainties to Identify Outliers}
The BNN with the outlier removal method discussed in \S 3.1 stands out as the overall best performing model for the majority of photo-z performance metrics considered in this work, achieving the lowest percentage of outliers, catastrophic outliers, and RMS error. The outlier removal method described in \S 3.1 is visualized in Figs. \ref{fig4},\ref{fig5},\ref{fig6}, \ref{fig8}, and \ref{fig9}. Notably, Fig. \ref{fig5} shows the performance of the NN and BNN with respect to LSST photo-z science requirements. The utilization of the photo-z uncertainties produced by the BNN to remove poor predictions significantly reduces RMS error. The BNN satisfies the LSST photo-z science requirements with respect to RMS error and $3 \sigma$ outlier fraction across $0 < z < 2.5$, however the bias requirement is only partially met in the range $0.3 < z < 1.2$. We note that the BNN bias deviation from the acceptable range is confluent with the drop-off of the galaxy population in the N(z) distribution in Fig. \ref{fig1}.

\subsection{Bayesian Neural Network Photo-z Uncertainty Estimates}
\begin{figure*}[!tbh]
  \centering
  \begin{minipage}[b]{0.45\textwidth}
    \includegraphics[width=\textwidth]{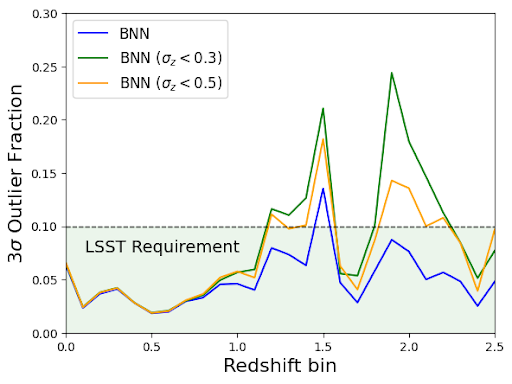}\\
    \includegraphics[width=\textwidth]{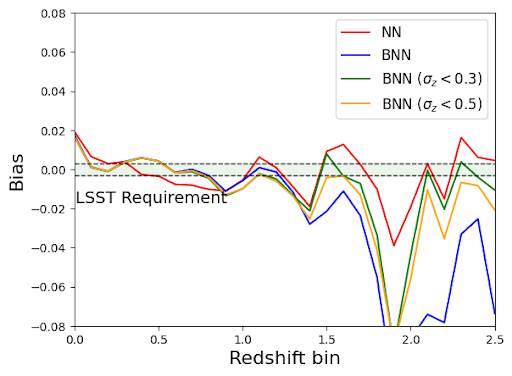}\\
  \end{minipage}
  \hfill
  \begin{minipage}[b]{0.45\textwidth}
    \includegraphics[width=\textwidth]{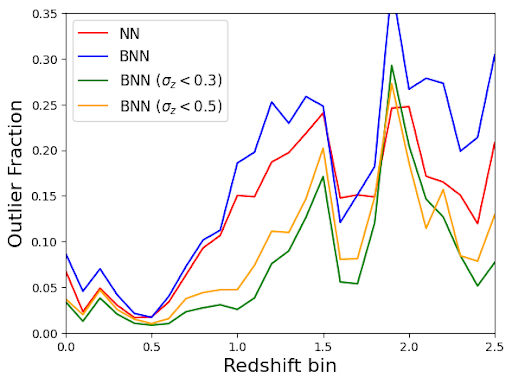}\\
    \includegraphics[width=\textwidth]{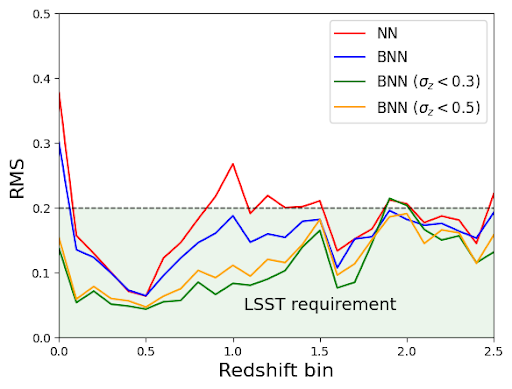}\\
  \end{minipage}
 \caption{BNN and NN performance with respect to LSST photo-z requirements. We note that the 3$\sigma$ outlier fraction can only be calculated with the BNN because the metric requires photo-z uncertainties so we additionally include the standard outlier fraction for the NN and BNN for comparison. The plots reflect results with $80\%$ of galaxies for training, $10\%$ for validation, and $10\%$ for evaluation. We include only those results in the redshift range $0 < z < 2.5$ because the $N(z)$ distribution of the data set degrades significantly at higher redshifts (see Fig. \ref{fig1}) and would likely significantly improve given sufficient training data.}
 \label{fig5}
\end{figure*}

\begin{figure}[!tbh]
\resizebox{\hsize}{!}
{\includegraphics{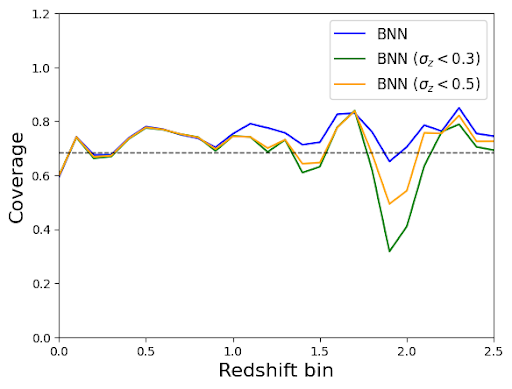}}
\label{z_rems}
\caption{The fraction of galaxies that have a spectro-z within their 68$\%$ confidence interval. Ideally, 68$\%$ of evaluated galaxies should have true spectro-zs within their 68$\%$ confidence interval. If more than 68$\%$ of evaluated galaxies have spectro-zs within their 68$\%$ confidence interval, the galaxies are considered `over-covered' because their photo-z uncertainties are too large. The same logic applies for `under-covered' galaxies. }
\label{fig6}
\end{figure}

\begin{figure*}
    \includegraphics[width=.33\textwidth]{f71.png}\hfill
    \includegraphics[width=.33\textwidth]{f72.png}\hfill
    \includegraphics[width=.33\textwidth]{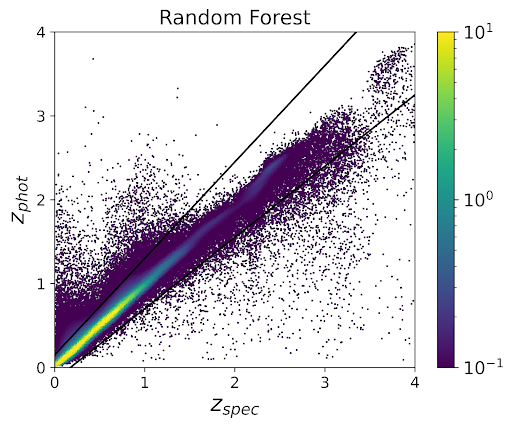}
    \\[\smallskipamount]
    \includegraphics[width=.33\textwidth]{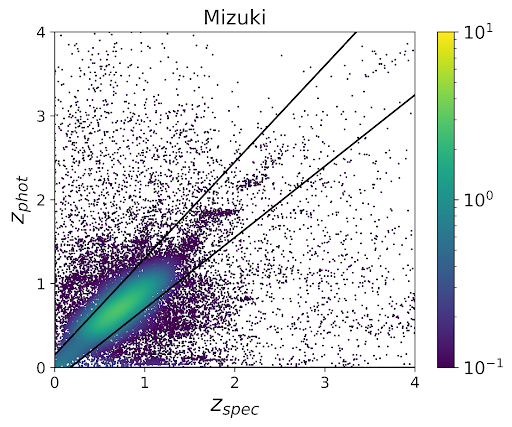}\hfill
    \includegraphics[width=.33\textwidth]{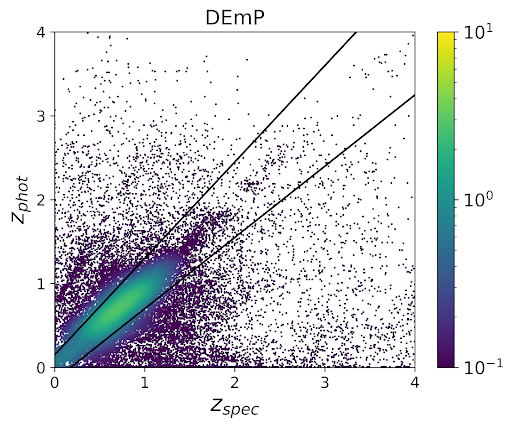}\hfill
    \includegraphics[width=.33\textwidth]{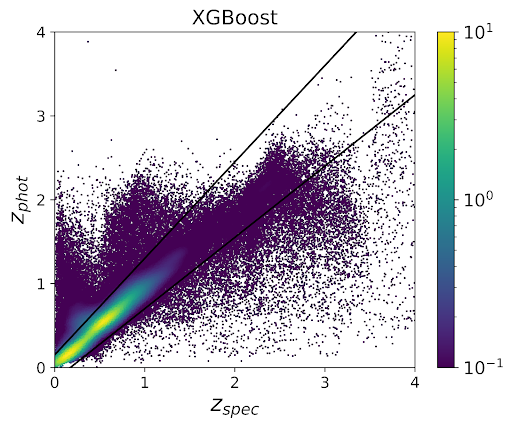}
        \\[\smallskipamount]
        \includegraphics[width=.33\textwidth]{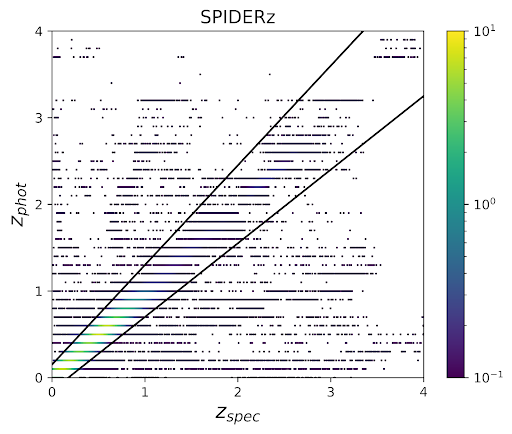}\hfill
    \caption{Visualization of predicted photo-zs versus measured spectroscopic redshifts by the models discussed in \S 2. The results of these determinations are quantified in Table 3. The colorbars indicate the density of evaluation data points as computed with a Gaussian kernel-density estimation.}\label{fig:foobar}
    \label{fig7}
\end{figure*}

\begin{figure*}[!tbh]
  \centering
  \begin{minipage}[b]{0.45\textwidth}
    \includegraphics[width=\textwidth]{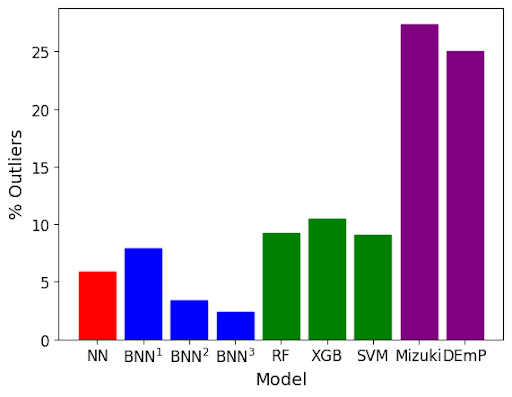}
  \end{minipage}
  \hfill
  \begin{minipage}[b]{0.45\textwidth}
    \includegraphics[width=\textwidth]{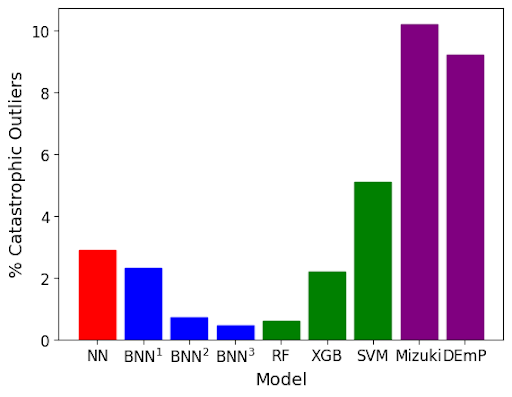}
  \end{minipage}
 \caption{Comparison of the percentage of outliers (Eqn 1) and catastrophic outliers (Eqn 2) achieved with each model. BNN$^1$ refers to the default BNN results with no galaxies removed based on a $z_{\sigma}$ criteria. BNN$^2$ and BNN$^3$ refer to results obtained after removing all galaxies from the evaluation set containing photo-z uncertainties greater than 0.5 and 0.3, respectively. We use the data discussed in \S \ref{data} to train and evaluate a NN, BNN, a SVM SPIDERz \citep{jones_analysis_2017}, a random forest \citep{breiman_random_2001}, and a gradient boosting model XGBoost \citep{chen_xgboost_2016}. We also include a comparison to the template-fitting model, Mizuki, and empirical method, DEmP \citep{ hsieh_estimating_2014}, that were evaluated on a larger, overlapping data set in \citep{nishizawa_photometric_2020}. To form a comparison to Mizuki and DEmP in this work, we crossmatched the larger data set with the object IDs of our data discussed in \S \ref{data} to obtain a pre-evaluated sample of ~60 thousand galaxies.}
 \label{fig8}
\end{figure*}

We find that the BNN produces accurate uncertainties as defined by the probabilistic metrics. The quality of the uncertainties produced by the BNN are visualized in Figs \ref{fig5}, \ref{fig6}, and \ref{fig10}. The BNN 3$\sigma$ outlier fraction is shown in Fig. \ref{fig5}, which indicates that uncertainties are generally well-estimated on average across the redshift range $0 <z < 2.5$. It is notable that the BNN performs best with respect to the $\sigma$ outlier fraction when no galaxies with large uncertainties are removed. The PIT histogram produced for a sample determination with the BNN is shown in Fig. \ref{fig10}. The PIT histogram is generally flat, as is desired, however the slight bump in the middle indicates that the photo-z PDFs tend to be overly broad. For a comparison of PITs produced by other probabilistic photo-z methods (performed on different data) see \citet{schmidt_evaluation_2020}. The BNN uncertainty coverage of the sample is provided in Fig. \ref{fig6}, showing acceptable agreement with the target 68$\%$ confidence interval up to the target redshift interval for weak lensing applications $0.3 < z < 1.5$, indicating the uncertainties of photo-z estimates for this galaxy population are accurately defined. 

The results from evaluating the NN and BNN models on the evaluation set are available at \url{https://zenodo.org/doi/10.5281/zenodo.10145347}. 

\subsection{Investigating the effect of non-representative training data}

The distribution of brightness of the training sample is peaked at brighter magnitudes compared to the full HSC photometric sample because of the need for spectroscopy. We investigated how this bias might affect our results by re-sampling the testing dataset to have a magnitude distribution closer to original HSC dataset. We find that the performance on this re-sample is similar or slightly worst by about 1 to 2\% depending on the metric. See Appendix A for more details. 



\begin{table*} 
\centering
\caption{Comparison of the performance results with each model discussed in \S 2. We use the data discussed in \S \ref{data} to train and evaluate a NN, BNN, a SVM SPIDERz \citep{jones_analysis_2017}, a random forest \citep{breiman_random_2001}, and a gradient boosting model XGBoost \citep{chen_xgboost_2016}. We also include a comparison to the template-fitting model, Mizuki, and empirical method, DEmP \citep{ hsieh_estimating_2014}, that were evaluated on a larger, overlapping data set in \citep{nishizawa_photometric_2020}. To form a comparison to Mizuki and DEmP in this work, we crossmatched the larger data set with the object IDs of our data discussed in \S \ref{data} to obtain a pre-evaluated sample of ~60 thousand galaxies.}
\centering
\begin{tabular}{lrrrrrrrrrrrr}
\hline
\hline
Network & $O$ & $O_c$ & $O_b$ & RMS & $|b|$ & Scatter & $L(\Delta z)$ \\
\hline
BNN & 0.079 & 0.023 &0.023	&0.174	&0.013	&0.026	&0.105
\\
BNN (${\sigma}_z < 0.5$) & 0.034&	0.0071&	0.025&	0.0854&	0.002&	0.029&	0.066
 \\
BNN (${\sigma}_z < 0.3$) & 0.0236&	0.0045&	0.017&	0.0738&	0.002&	0.022&	0.056
 \\
NN & 0.059	&0.029	&-	&0.174&	0.0001&	0.026&	0.089
 \\

\hline
Mizuki  & 0.274 & 0.102 & - & 0.307 & 0.011 & 0.055 & 0.289 \\
DEmP  & 0.250 & 0.092 & - & 0.277 & 0.003 & 0.040 & 0.258\\
\hline
RF  & 0.092 & 0.006 & - & 0.088 & 0.001 & 0.012 & 0.065\\
XGBoost & 0.105 & 0.022& - & 0.149 & 0.002 & 0.033 & 0.144\\
SPIDERz & 0.090 & 0.051 & - & 0.199 & 0.002 & 0.044 & 0.135\\
\hline
LSST Req. & - & -& - & $<$ 0.2 & $<$ 0.003 & $<$ -  & - \\
\end{tabular}
\label{tab:results}
\end{table*}

\section{Discussion}
Future large-scale astronomical surveys will provide high quality observations of billions of celestial objects that will be used to investigate the mysterious and unknown nature of dark matter and dark energy. LSST will play a crucial part in this investigation; we model our analysis here with respect to the photo-z science requirements provided by the LSST team. Bayesian Neural Networks have been used in the past for photo-z estimation \citep[e.g.][]{zhou_photometric_2022, schuldt_photometric_2020, jones_photometric_2021}, however this is the first BNN (following our prototype in \citet{jones_photometric_2022}) applied to photometry observations of a representative dataset similar to what we will obtain from future large scale surveys like LSST. This work is among the first that evaluates the photo-z model performance with respect to the LSST science requirements as a function of redshift. 

The BNN largely satisfies LSST science requirements in the redshift range of interest for LSST weak lensing surveys ($0.3 < z < 1.5$), and outperforms alternative models on the same data, however the BNN does not fully satisfy the bias requirement. We believe the BNN model can be further optimized for these requirements. Compared to the NN model, the BNN has the advantage of producing uncertainties for each prediction, which are both required for precision cosmology and can be used to eliminate galaxies with large uncertainties from the data sample. We note that both the NN and BNN models generally perform worse at higher redshifts, which is due in large part to the reduced signal to noise for distant dim sources and also the disproportionate number of high redshift sources (z $>$ 2.5) compared to low redshift sources (see Fig. \ref{fig1} and also the discussion in \citet{wyatt_outlier_2020}). 

The BNN model introduced here is an improved version of the model we introduced in a previous work \citep{jones_photometric_2022}. The uncertainty estimates produced in the BNN model discussed in this work are significantly improved from the previous model -- due in large part to optimizing the learning rate during training and modifying the network architecture. The previous model network contained four variational layers, which we adjusted to contain three dense layers and one variational layer. We find that the coverage for the architecture with all variational layers produces coverage that is generally 10\% larger than expectation (uncertainties too large). Using a single variational now produces more accurate coverage.

The BNN uncertainty coverage of the sample provided in Fig. \ref{fig10} shows excellent agreement with the target 68$\%$ confidence interval up to $z = 1.5$, indicating the uncertainties of photo-z estimates for this galaxy population are accurately defined. Beyond $z = 1.5$, the coverage oscillates around the target $\%68$ level. A likely explanation for the reduced quality of galaxy uncertainties beyond $z = 1.5$ is the lack of data samples at this redshift range (see Fig. \ref{fig1}) compared to lower redshifts. Another possible factor affecting photo-z uncertainties may result from a disparity between the complexity present in the band magnitudes compared to the BNN model; we use five photometric band fluxes paired with a single spectroscopic redshift per galaxy for training. In a future work we will apply galaxy photometric images to a Bayesian convolutional neural network, which is likely to contribute more useful information than the five photometric measurements per galaxy.  

Another benefit of using a BNN for photo-z estimation is the use of the photo-z uncertainty $z_{\sigma}$ to preemptively flag photo-z predictions with high uncertainties as potential poor predictions. An acceptable balance needs to be achieved between the number of galaxies correctly flagged as poor predictions versus the number of non-outlier galaxies removed for a given $z_{\sigma}$ cutoff value. With the data used in this work we find a significant reduction in the number of catastrophic outliers and outliers can be achieved by sacrificing a relatively small number of non-outlier predictions; for example, we find that by removing all galaxies in the evaluation sample with a photo-z uncertainty ${\sigma}_z > 0.3$, the RMS error was reduced by 57.6$\%$, outliers were reduced by 70.1$\%$, and catastrophic outliers were reduced by $80.43\%$ -- at the cost of removing only 11$\%$ of the evaluation set.

\section{Conclusion}

In preparation of the coming influx of data from large scale surveys like LSST, it is important to prepare photo-z estimation models in advance. Such models must provide both accurate photo-z predictions and reliable photo-z uncertainties, which are required for using photo-z predictions in subsequent cosmological analyses. The quality of photo-z models should be assessed using data that is representative of data from future large scale surveys, and principally evaluated using the scientific requirements provided by those surveys.  

This work introduces a BNN model for photometric redshift estimation. We apply the BNN to data from the Hyper Suprime Cam survey, which is designed to reflect the data we will soon receive from large scale surveys such as LSST. We evaluate the BNN with respect to LSST science requirements and compare the results to alternative photo-z estimation tools including a fully connected neural network, random forest, support vector machine \citep{jones_analysis_2017}, XGBoost, Mizuki, and DEmp \citep{aihara_first_2018, aihara_second_2019}. We find that the BNN meets two of the three LSST photo-z requirements in the redshift range considered for weak lensing cosmological probes $(0.3 < z < 1.5)$ and provides superior photo-z estimations to the other models. 

A key attribute of the BNN model is the production of photo-z uncertainties, which are needed for using photo-z results in cosmological analyses. We find that the BNN produces accurate uncertainties. Using a coverage test, we find excellent agreement with expectation -- 68.5$\%$ of galaxies between $0 < 2.5$ have 1-$\sigma$ uncertainties that cover the spectroscopic value. In addition, the BNN photo-z uncertainties can be used to flag likely outlier or catastrophic outlier estimates with high success. 

This analysis is subject to the potential sources of bias that affect most photometric redshift estimation studies. For example, spectroscopic redshift observations are biased toward high luminosity galaxies, particularly at higher redshift ranges ($z > 1.5$), which may not be fully representative of galaxy populations at a specific redshift range. Another source of bias in this analysis is the underrepresented galaxy population in the N(z) distribution beyond z = 1.5. Both of these potential sources of bias can be alleviated with improved spectroscopic samples in future galaxy surveys.

We will continue this analysis by applying the BNN method to galaxy images via a Bayesian convolutional neural network in a forthcoming paper.

We are grateful for the financial support for this work from the Sloan Foundation.


\begin{figure}[!h]
\resizebox{\hsize}{!}
{\includegraphics{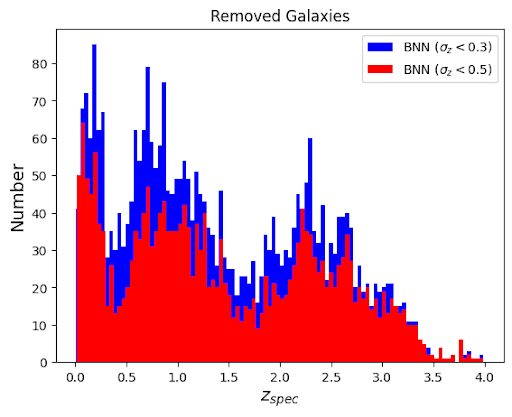}}
\label{z_rems}
\caption{Histogram of photo-z uncertainties produced by the BNN that exceed 0.3 and 0.5. By removing all galaxies in the evaluation sample with a photo-z uncertainty ${\sigma}_z < 0.3$, outliers were reduced by 70.1$\%$, and catastrophic outliers were reduced by $80.43\%$ -- at the cost of removing  11$\%$ of the evaluation set. Using a photo-z uncertainty cutoff of 0.5 reduces the number of outliers by $70.1\%$ and catastrophic outliers by $67.8\%$ at the cost of removing $7.67\%$ of the evaluation set.}
\label{fig9}
\end{figure}

\begin{figure}[!tbh]
\resizebox{\hsize}{!}
{\includegraphics{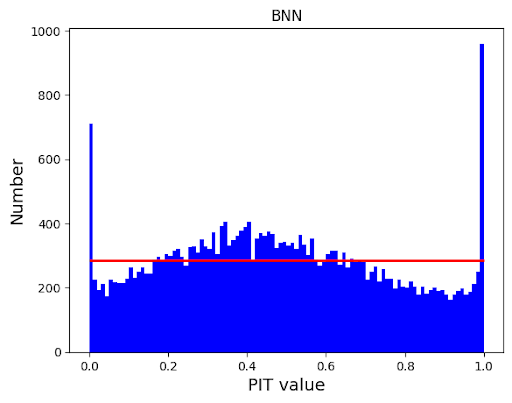}}
\label{z_rems}
\caption{PIT histogram of the photo-z PDF produced by the Bayesian Neural Network. The red horizontal line indicates the ideal PIT histogram distribution: if the PIT histogram peaks at the center, the photo-z PDFs are generally too broad, and if the PIT histogram peaks at high and low PIT values, the PDF samples are too narrow or contain a large amount of catastrophic outliers.}
\label{fig10}
\end{figure}

\bibliography{ms.bib}{}
\bibliographystyle{aasjournal}

\appendix{
\section{Addressing potential biases in the dataset}
Because our selection of data for training and evaluation relied on only those galaxies for which spectroscopic redshifts are available, the magnitude distribution is biased compared to a the bulk photometric sample from HSC (see Fig. 11). In order to address this, we have performed an additional analysis with the NN and BNN models using a re-sampled testing set that mimics the magnitude distribution of the bulk HSC photometry sample (Fig. \ref{fig11}). We use the g-band to re-sample our testing dataset to reproduce the overall HSC g-band distribution.  Since the color of the resampled dataset is not enforced, the resampled distribution in the \textit{rizy} filters are slightly different than the main HSC distribution. However, these distributions are all much closer than the initial spectrosopic sample. The re-sampling process reduced our testing dataset size from 28,640 to 8,517 $--$ a $70.3\%$ decrease.  The distribution of redshifts for the resampled testing data is similar to the original (Fig. \ref{fig12}).



\begin{figure*}[!tbh]
  \centering
  \begin{minipage}[b]{0.49\textwidth}
    \includegraphics[width=\textwidth]{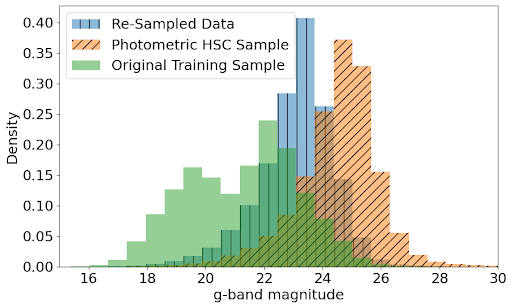}\\
    \includegraphics[width=\textwidth]{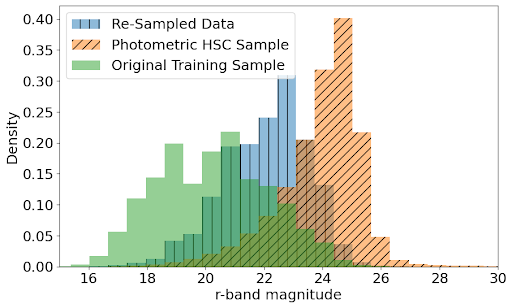}\\
    \includegraphics[width=\textwidth]{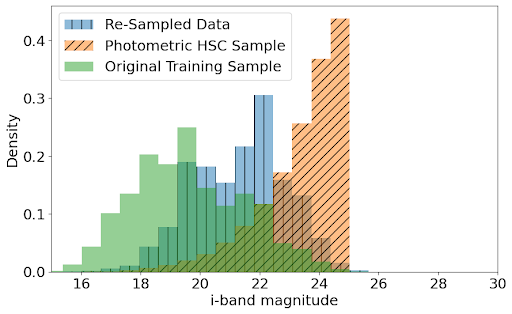}\\
  \end{minipage}
  \hfill
  \begin{minipage}[b]{0.49\textwidth}
    \includegraphics[width=\textwidth]{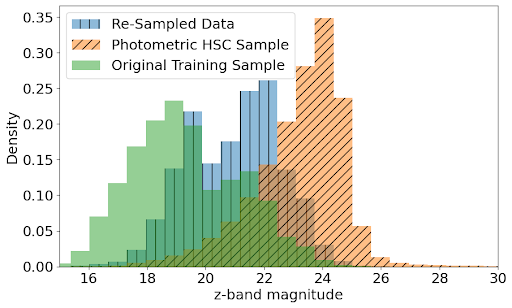}\\
    \includegraphics[width=\textwidth]{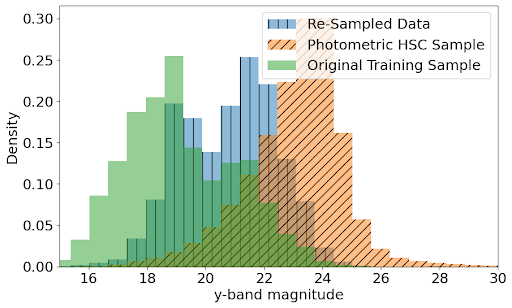}\\
    \end{minipage}
 \caption{Visualisation of the grizy bands before and after the data is re-sampled to approximate the bulk HSC photometry sample.}
 \label{fig11}
\end{figure*}



Overall, the model does not perform significantly different on the resampled testing dataset (Fig. \ref{fig13},\ref{fig14},\ref{fig15},\ref{fig16},\ref{fig17}). Overall, the resampled testing data performs about $1-2\%$ worse than the original testing dataset. The coverage of the re-sampled data is not signficantly affected either (Fig. \ref{fig14}). In addition, the PIT histogram (Fig. \ref{fig16}) indicates that the photo-z PDFs produced in the re-sampled testing set closely resemble the photo-z PDFs produced in the original evaluation sample shown in in Fig. \ref{fig9}. Table \ref{tab:resampled} quantifies the performance metrics shown in Table \ref{tab:results}. 


\begin{figure}[!tbh]
\resizebox{\hsize}{!}
{\includegraphics{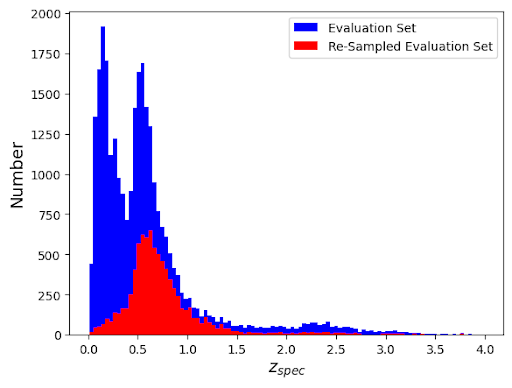}}
\label{z_rems}
\caption{N(z) distributions of the original evaluation set discussed in \S \ref{data} and the re-sampled evaluation set.}
\label{fig12}
\end{figure}

\begin{figure*}[!tbh]
  \centering
  \begin{minipage}[b]{0.45\textwidth}
    \includegraphics[width=\textwidth]{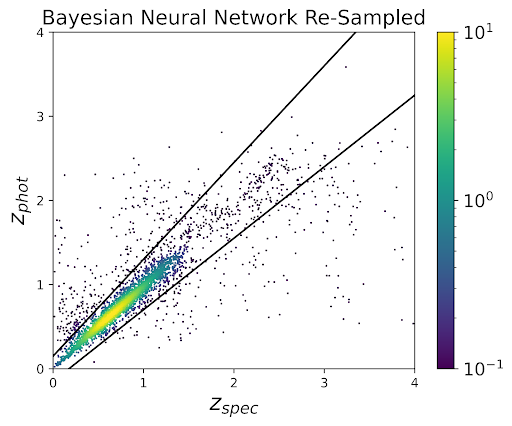}\\
  \end{minipage}
  \hfill
  \begin{minipage}[b]{0.45\textwidth}
    \includegraphics[width=\textwidth]{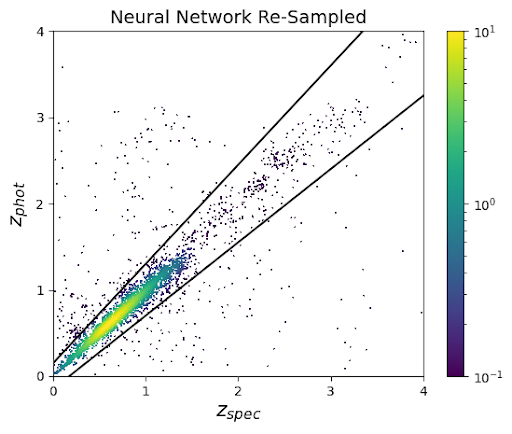}\\
  \end{minipage}
\caption{Visualization of the NN and BNN results using an evaluation set that is re-sampled to more closely approximate the bulk HSC photometry. The models are trained on the original data discussed in \S \ref{data}.}
\label{fig13}
\end{figure*}

\begin{figure}[!tbh]
\resizebox{\hsize}{!}
{\includegraphics{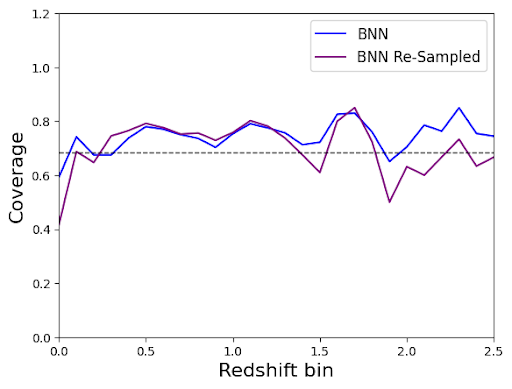}}
\label{z_rems}
\caption{Comparison of the photo-z uncertainty coverage present in the original evaluation set compared to the re-sampled evaluation sample. Coverage is defined as the fraction of galaxies that have a spectro-z within their 68$\%$ confidence interval. Ideally, 68$\%$ of evaluated galaxies should have true spectro-zs within their 68$\%$ confidence interval. If more than 68$\%$ of evaluated galaxies have spectro-zs within their 68$\%$ confidence interval, the galaxies are considered `over-covered' because their photo-z uncertainties are too large. The same logic applies for `under-covered' galaxies. }
\label{fig14}
\end{figure}

\begin{figure*}[!tbh]
  \centering
  \begin{minipage}[b]{0.45\textwidth}
    \includegraphics[width=\textwidth]{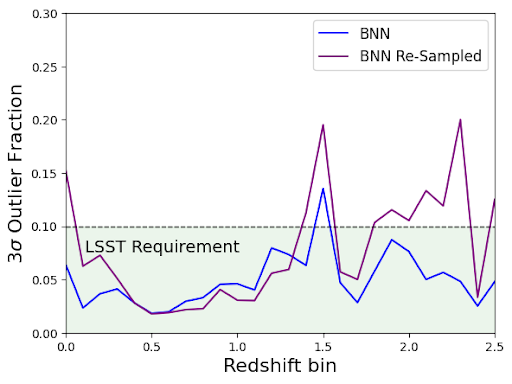}\\
    \includegraphics[width=\textwidth]{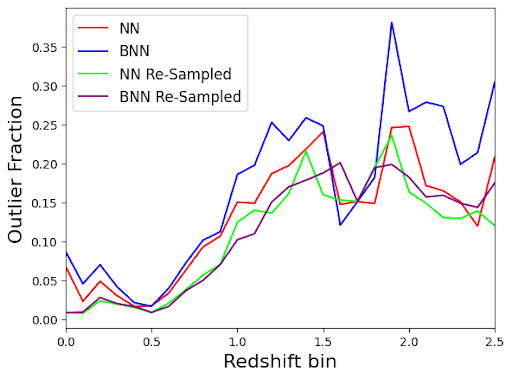}\\
  \end{minipage}
  \hfill
  \begin{minipage}[b]{0.45\textwidth}
    \includegraphics[width=\textwidth]{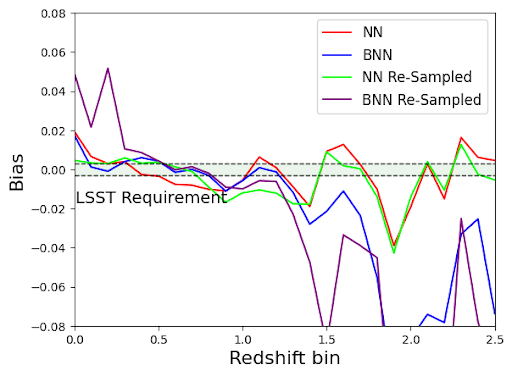}\\
    \includegraphics[width=\textwidth]{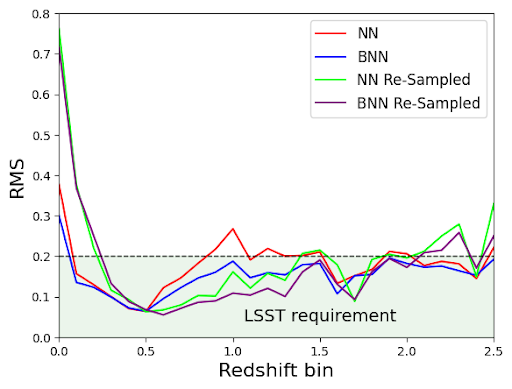}\\
  \end{minipage}
 \caption{BNN and NN performance with respect to LSST photo-z requirements using an evaluation set with photometry that is re-sampled to approximate the bulk HSC photometry. We note that the 3$\sigma$ outlier fraction can only be calculated with the BNN because the metric requires photo-z uncertainties so we additionally include the standard outlier fraction for the NN and BNN for comparison. The plots reflect results with $80\%$ of galaxies for training, $10\%$ for validation, and $10\%$ for evaluation. We include only those results in the redshift range $0 < z < 2.5$ because the $N(z)$ distribution of the data set degrades significantly at higher redshifts (see Fig. \ref{fig1}) and would likely significantly improve given sufficient training data.}
 \label{fig15}
\end{figure*}

\begin{figure}[!tbh]
\resizebox{\hsize}{!}
{\includegraphics{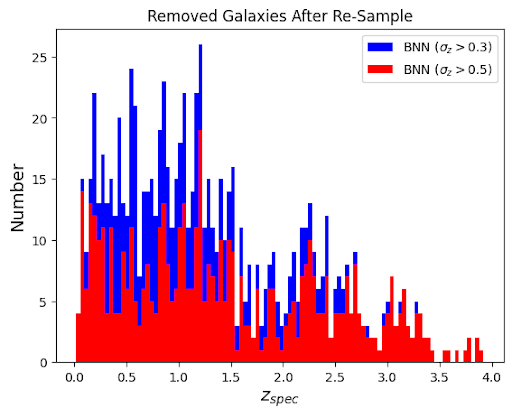}}
\label{z_rems}
\caption{Histogram of photo-z uncertainties produced by the BNN that exceed 0.3 and 0.5 using the re-sampled dataset. By removing all galaxies in the evaluation sample with a photo-z uncertainty ${\sigma}_z < 0.3$, outliers were reduced by 70.1$\%$, and catastrophic outliers were reduced by $80.43\%$ -- at the cost of removing  11$\%$ of the evaluation set. Using a photo-z uncertainty cutoff of 0.5 reduces the number of outliers by $70.1\%$ and catastrophic outliers by $67.8\%$ at the cost of removing $7.67\%$ of the evaluation set.}
\label{fig16}
\end{figure}

\begin{figure}[!tbh]
\resizebox{\hsize}{!}
{\includegraphics{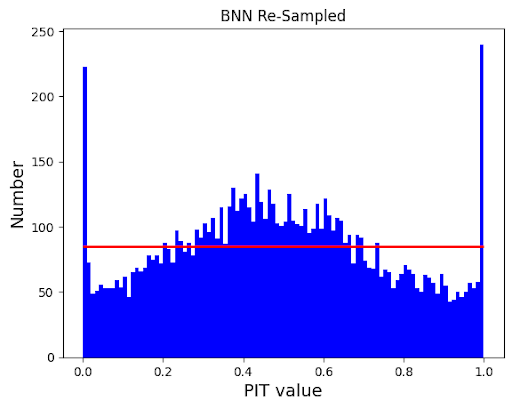}}
\label{z_rems}
\caption{PIT histogram of the photo-z PDF produced by the Bayesian Neural Network using an evaluation set with photometry that is re-sampled to approximate the HSC bulk photometry. The red horizontal line indicates the ideal PIT histogram distribution: if the PIT histogram peaks at the center, the photo-z PDFs are generally too broad, and if the PIT histogram peaks at high and low PIT values, the PDF samples are too narrow or contain a large amount of catastrophic outliers.}
\label{fig17}
\end{figure}

\begin{table}[h]
\centering
\begin{tabular}{lccccccc}
\hline
Network & $O$ & $O_c$ & $O_b$ & RMS & $|b|$ & Scatter & $L(\Delta z)$ \\ \hline
BNN         & 0.079          & 0.023               & 0.0233                   & 0.174        & 0.013      & 0.026           & 0.1054        \\
BNN re-sampled  & 0.08           & 0.017                & 0.026                   & 0.134        & 0.0047      & 0.028           & 0.1082        \\
NN          & 0.059            & 0.029                & 0.174                    & 0.0001       & 0.026        & 0.089            & 0.089         \\
NN re-sampled   & 0.067           & 0.021               & 0.14                   & -0.0029      & 0.027        & 0.097           & 0.097        \\
 \hline
\end{tabular}
\caption{Comparison of the performance results with the NN and BNN with the original evaluation data set discussed in \S \ref{data} and the re-sampled evaluation set to approximate the bulk HSC photometry. We use the data discussed in \S \ref{data} to train. The re-scaling process reduces the initial evaluation set size from 28,640 to 8,517 $--$ a $70.3\%$ decrease.}
\label{tab:resampled}
\end{table}
\end{document}